\documentclass{agujournal}
\journalname{Journal of Geophysical Research: Space Physics}

\usepackage{lineno}
\usepackage{amsmath,amssymb,graphicx,url,color}

\begin{document}

\title{Recurrence based quantification of dynamical complexity in the Earth's magnetosphere at geospace storm timescales}
\authors{Reik V. Donner\affil{1},
Georgios Balasis\affil{2},
Veronika Stolbova\affil{1,3,4,5},
Marina Georgiou\affil{2,6,7},
Marc Wiedermann\affil{1,3},
and J\"urgen Kurths\affil{1,3,4,7}}

\affiliation{1}{Research Domain IV -- Transdisciplinary Concepts and Methods, Potsdam Institute for Climate Impact Research, Potsdam, Germany}
\affiliation{2}{Institute for Astronomy, Astrophysics, Space Applications and Remote Sensing, National Observatory of Athens, Penteli, Athens, Greece}
\affiliation{3}{Department of Physics, Humboldt University, Berlin, Germany}
\affiliation{4}{Department of Banking and Finance, University of Zurich, Switzerland}
\affiliation{5}{Department of Control Theory, Nizhny Novgorod State University, Nizhny Novgorod, Russia}
\affiliation{6}{Department of Physics, Section of Astrophysics, Astronomy and Mechanics, National and Kapodistrian University of Athens, Zografos, Athens, Greece}
\affiliation{7}{Mullard Space Science Laboratory, Department of Space \& Climate Physics, University College London, Dorking, Surrey, United Kingdom}
\affiliation{8}{Institute for Complex Systems and Mathematical Biology, University of Aberdeen, Aberdeen, United Kingdom}

\correspondingauthor{Reik V. Donner}{reik.donner@pik-potsdam.de}

\begin{keypoints}
\item Dst index and three solar wind variables are characterized by nonlinear characteristics based on recurrence plots.
\item Recurrence characteristics reveal different mean levels and marked temporal changes of dynamical complexity in the four variables.
\item Different geospace storm periods are accompanied by different dynamical complexity profiles in the interplanetary forcing.
\end{keypoints}

\begin{abstract}
Magnetic storms are the most prominent global manifestations of out-of-equilibrium magnetospheric dynamics. Investigating the dynamical complexity exhibited by geomagnetic observables can provide valuable insights into relevant physical processes as well as temporal scales associated with this phenomenon. In this work, we utilize several innovative data analysis techniques enabling a quantitative nonlinear analysis of the nonstationary behavior of the disturbance storm time (Dst) index together with some of the main drivers of its temporal variability, the $VB_{South}$ electric field component, the vertical component of the interplanetary magnetic field, $B_z$, and the dynamic pressure of the solar wind, $P_{dyn}$. Using recurrence quantification analysis (RQA) and recurrence network analysis (RNA), we obtain several complementary complexity measures that serve as markers of different physical processes underlying quiet and storm time magnetospheric dynamics. Specifically, our approach discriminates the magnetospheric activity in response to external (solar wind) forcing from primarily internal variability and provides a physically meaningful classification of magnetic storm periods based on observations made at the Earth's surface. In this regard, the proposed methodology could provide a relevant step towards future improved space weather and magnetic storm forecasts.
\end{abstract}


\section{Introduction}

Geospace magnetic storms are major perturbations of the Earth's magnetic field that are initiated by enormous bursts of plasma erupting from the solar corona. In addition to coronal mass ejections (CMEs), high-speed solar streams emanating from coronal holes provide solar wind structures that create favorable conditions for the development of magnetic storms. The ejection of highly energetic charged particles onto a trajectory intersecting with the Earth's orbit can have severe impacts on the Earth's magnetosphere \citep{Bothmer2007,Richardson2012}. Similar to other extreme events in nature, the resulting perturbations of the geomagnetic field can vary remarkably in both magnitude and duration. However, unlike many other natural hazards, they commonly manifest themselves in simultaneous effects all around the globe.

The mechanism underlying these large-scale perturbations of the Earth's magnetic field is closely related with mass, momentum and energy input provided by the solar wind that is stored in the magnetotail -- if not dissipated. Due to this continuous input by the highly dynamic solar wind, the magnetosphere is always far from equilibrium \citep{Consolini2008}. When a critical threshold is reached, the magnetospheric system may be reconfigured through a sequence of energy and stress accumulating processes \citep{Klimas1997,Klimas1998,Klimas2005,Baker2007}. During major magnetic storms, charged particles confined in the Earth's radiation belts are accelerated to high energies and the intensification of electric current systems results in characteristic disturbances of the geomagnetic field \citep{Baker2005,Daglis2008}. The response of the magnetosphere to the external forcing by the solar wind is in general not simply proportional to the input, and changes are episodic and abrupt, rather than slow and gradual. 

This distinct behavior motivated the description of the Earth's magnetosphere as a complex system composed of several nonlinearly coupled sub-systems, within which multiple interconnected processes act on a wide range of spatial and temporal scales \citep[][and references therein]{Chang1992,Klimas1996,Watkins2001,Consolini2002,Valdivia2005}. \citet{Vassiliadis1990} provided evidence of large-scale coherence in magnetosphere dynamics manifested as low-dimensional chaos in time series of auroral electrojet (AE) index measurements. Following upon these initial findings, subsequent studies have been based on a variety of complementary concepts from nonlinear dynamics and complex systems science to derive in-depth knowledge on the magnetosphere's response to the solar wind forcing. Among other approaches, the nonlinearity of magnetospheric dynamics has been studied using nonlinear filters \citep{Vassiliadis1995,Weigel2003}, explicitly accounting for the magnetosphere being a non-autonomous, driven system and contributing to a more accurate and efficient prediction of imminent magnetic storms \citep{Valdivia1996}. 

Building upon the current understanding of magnetic field fluctuations at the Earth's surface and in the surrounding space, recent studies drew the picture of the magnetosphere as a hierarchically organized multiscale system based on power law dependencies identified in time series of geomagnetic activity indices \citep{Takalo1994,Consolini1997,Chapman1998,Uritsky1998,Watkins2001,Wanliss2004,Wei2004}. For instance, the adoption of a phase transition approach \citep{Shao2003,Sharma2006} revealed a close connection between global coherence and scale invariance of the magnetosphere's behavior. Specifically, \citet{Sitnov2001} suggested that while the multiscale activity during substorms resembles second order phase transitions, the largest substorm avalanches exhibit common features of first order nonequilibrium transitions. Moreover, \citet{Balasis2006} demonstrated the existence of two different regimes in the magnetospheric dynamics associated with the pre-storm activity and magnetic storms, respectively, a picture which is compatible with the occurrence of a phase transition. Low-dimensional dynamics, self-organized criticality \citep{Consolini1997,Uritsky1998,Chapman1998,Uritsky2006} and phase transitions offer different perspectives on geomagnetic activity, all of which need to be taken into account to obtain a coherent global picture of the underlying dynamical processes. However, when considered individually, each of these approaches has its intrinsic limitations that are inherent to the specific methodology and respectively taken viewpoint on specific aspects of nonlinear dynamics. 

This paper aims to offer an additional viewpoint on nonlinear magnetospheric variability that has not yet been systematically addressed in previous studies. Specifically, we utilize a set of complementary measures characterizing the dynamical complexity of time series provided by the powerful tools of recurrence analysis \citep{Marwan2007}. The main idea behind this approach is that most physical processes lead to recurrences of previous states or sequences thereof, meaning that a system's current dynamical state has some close analog in its past dynamics, both of which are separated by some period with different system properties. Recent developments in dynamical systems theory have provided evidence that such a behavior is a generic property of both deterministic and stochastic dynamics \citep{Marwan2007,Marwan2014}. Among the existing nonlinear time series analysis approaches based on the evaluation of such recurrences, recurrence quantification analysis (RQA) \citep{Marwan2007} and recurrence network analysis (RNA) \citep{Donner2011} have already proven their potential for tracing time-varying dynamical complexity in a wide variety of different fields \citep{Bastos2011,Zbilut2002,Schinkel2009,Donges2011,Donges2011PNAS}. 

In this work, we apply RQA and RNA to trace dynamical complexity variations in the Earth's magnetic field that have distinct signatures in geomagnetic activity indices. We start by investigating hourly recordings of the disturbance storm time (Dst) index, one of the most intensively studied indicators for magnetospheric variability on time scales between hours and weeks (which integrates over any higher-frequency magnetic field fluctuations and is therefore less prone to exhibit bursty short-term dynamics than other conceptually related geomagnetic indices like Sym-H \citep{Wanliss2006}). Specifically, the Dst index is defined as the average change of the horizontal component of the Earth's magnetic field recorded at four mid-latitude magnetic observatories and therefore provides an integral picture of the overall state of the magnetosphere. Although Dst alone cannot represent the full complexity associated with geomagnetic storms (i.e., phenomena with very specific electromagnetic and particle signatures in different regions of the inner magnetosphere, which can be more directly observed by in situ measurements like those provided by the Van Allen probes) and thus cannot be used as a unique classifier for magnetic storms, its continuous availability makes it a useful proxy for studying long-term variations of the Earth's magnetosphere.

Notably, the geomagnetic field component represented by the Dst index is determined by both, solar wind forcing and internal magnetospheric processes. By comparing temporal changes in the dynamical complexity of Dst index fluctuations with those of variables associated with the driving processes, in this work we aim to disentangle the dynamical signatures originating from internal magnetospheric complexity from the additional complexity enforced by the solar wind. Specifically, we present the results of recurrence analysis for time variations of the $VB_{South}$ electric field component acting as a coupling function between external (solar wind) and geomagnetic field, as well as for the vertical component of the interplanetary magnetic field (IMF) $B_z$ and the solar wind dynamic pressure $P_{dyn}$, potentially allowing us to differentiate the internal magnetospheric variability from its response to external forcing. By means of recurrence analysis, we are able to resolve subtle aspects of magnetospheric dynamics resulting from nonlinear interactions between subsystems, which have been hidden to previously employed methods. The properties explored here have not yet been captured by any other linear or nonlinear method of time series analysis previously applied for studying these four specific variables.

This paper is structured as follows: in Section~\ref{sec:data}, we describe in detail the observational data used in this study, while the recurrence analysis methods employed to trace different aspects of the time-dependent dynamical complexity of the coupled solar wind-magnetosphere system are discussed in Section~\ref{sec:method}. The resulting mean dynamical complexity and its associated temporal variability as revealed by different recurrence based measures for the Dst index and solar wind parameters are detailed in Section~\ref{sec:performance}, while implications of our results are addressed in Section~\ref{sec:discussion}. A summary of our main findings is provided in the concluding Section~\ref{sec:conclusion}.

\section{Description of the data} \label{sec:data}

The solar cycle 23 (May 1996 to January 2008) has been characterized by a prolonged activity maximum that lasted from 2000 to 2003 and exhibited numerous strong solar eruptions followed by enhanced Earth's magnetospheric activity. In this work, we focus on one year of observations during this solar activity maximum from 1 January 2001 to 31 December 2001. This time period is of particular interest in the context of the present analysis, because it was marked by two unusually large CMEs on 29 March 2001 and 4 November 2001 that were followed by intense magnetic storms on 31 March 2001 and 6 November 2001, when the Dst index reached minimum values of $-387$~nT and $-292$~nT, respectively. These two magnetic storms marked the respective peaks of two time periods of intense magnetospheric disturbances that were separated by a phase of relatively low geomagnetic activity. Note that in general, a seasonal variation observed in the geomagnetic activity under extreme solar wind conditions during the solar cyle maximum can be explained by the Russell--McPherron effect \citep{Russell1973,Zhao2012}: Geomagnetic activity is much more intense around the spring equinox (when the IMF is directed towards the Sun) and around the fall equinox (when the direction of IMF points away of the Sun) than during the rest of the year.

\begin{figure}
\centering
\resizebox{0.5\textwidth}{!}{\includegraphics*{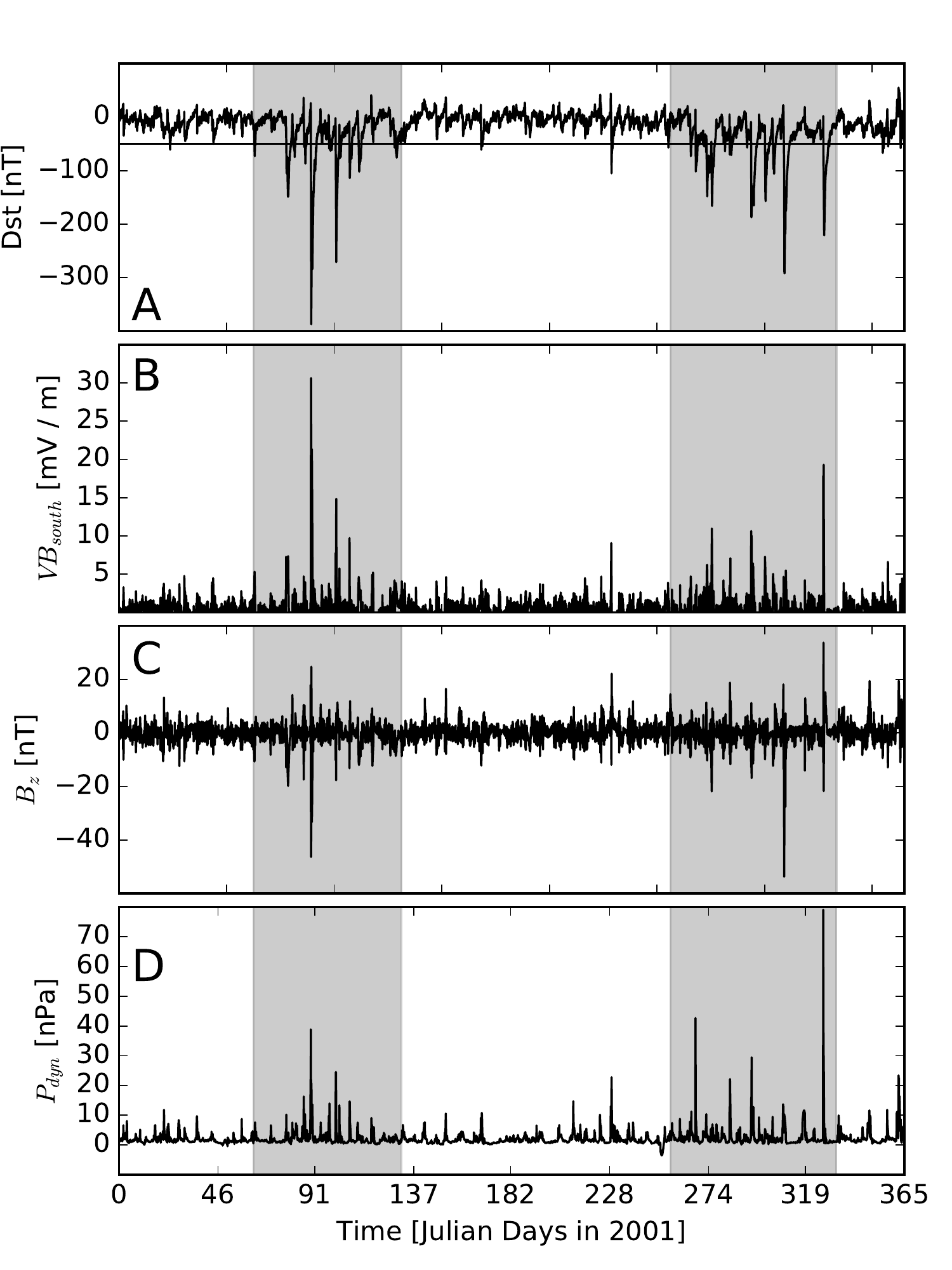}}
\caption{Time series of four variables characterizing the states of the Earth's magnetosphere and the solar wind: (A) Dst index, (B) $VB_{south}$, (C) $B_z$ and (D) $P_{dyn}$. All data have hourly time-resolution and cover the period from 1 January 2001 to 31 December 2001. Gray shaded areas indicate two periods (Julian days 65--110 and 285--330 of the year 2001) characterized by intense magnetic storm activity, whereas the other intervals are considered as periods of relative magnetospheric quiescence. The horizontal line in the upper panel corresponds to a value of Dst = -50~nT, which is commonly considered as a threshold for defining a magnetic storm.}
\label{fig:data}
\end{figure}

Figure~\ref{fig:data} shows the time series of the hourly Dst index for the entire year 2001 together with three key variables of the solar wind: the $VB_{south}$ component of the electric field, the $B_z$ magnetic field component and the dynamic pressure $P_{dyn}$. The Dst index measurements have been obtained from the World Data Center for Geomagnetism of the Kyoto University at \url{http://wdc.kugi.kyoto-u.ac.jp/index.html}, whereas the interplanetary data have been retrieved through the NASA space physics data facility OMNIWeb at \url{http://omniweb.gsfc.nasa.gov/}. In the case of $P_{dyn}$ a few missing values in the considered time series have been filled by employing a gap filling procedure based on singular spectrum analysis \citep{Kondrashov2006,Buttlar2014}.

According to the overall mean state of the coupled solar wind-magnetosphere system during time intervals of the order of weeks, our previous work has identified five distinct segments in these time series data based on the general geomagnetic activity level together with various associated dynamical characteristics \citep{Balasis2006, Balasis2008, Balasis2009, Balasis2011a,Donner2013}. The second and fourth segment (gray-shaded areas in Fig.~\ref{fig:data}) have been characterized by enhanced solar and magnetospheric activity and include the two aforementioned intense magnetic storms of March and November 2001. The remaining three time intervals correspond to a rather quiescent Earth magnetosphere. 

More specifically, our previous studies \citep{Balasis2006, Balasis2008, Balasis2009, Balasis2011a,Donner2013} -- mostly focusing on the Dst index alone -- have shown that the variability of the Earth's magnetosphere during these periods of geomagnetic activity and quiescence was characterized by two distinct patterns of dynamical organization in the Dst index: (i) periods with intense magnetic storms exhibit a markedly elevated degree of organization, representative of states of a ``disturbed'' magnetosphere, and (ii) typical non-storm periods, when the magnetosphere remained at ``normal'' states, with a lower degree of organization. It should be noted that this differentiation between distinct magnetospheric states has been based on the persistent versus anti-persistent character of the Dst index fluctuations \citep{Balasis2006} rather than the actual hourly Dst index values recorded within periods of different levels of geomagnetic activity.

\section{Recurrence analysis}\label{sec:method}

The nonlinear time series analysis methods employed in this study make use of the concept of phase space, an abstract metric space in which each possible system state is represented as a unique point. For deterministic dynamical systems, the state vectors at each point in time are fully determined by a minimal set of variables that govern the system's dynamical equations of motion (which are commonly not explicitly known).

\subsection{Phase space reconstruction by time-delay embedding}

In the case of univariate time series $\{x(t_i)\}_{i=1}^N$ -- presumably originating from a dissipative and (at least partially) deterministic dynamical system, as implicitly considered in the present work -- only one possible coordinate of the phase space is known explicitly. In such a case, phase space reconstruction by means of time delay embedding \citep{Packard1980,Mane1981,Takens1981} provides a widely applicable approach to qualitatively estimate the action of unobserved variables. A multivariate representation $X(t)$ in a new space, known as the reconstructed phase space or embedding space, is obtained from time-shifted replications of the original data: 
\begin{equation}
X(t)=(x(t),x(t-\tau),x(t-2\tau),...,x(t-(m-1)\tau)) \label{eq:embedding},
\end{equation}
\noindent
where the different coordinates of the reconstructed state vector $X(t)$  need to be sufficiently independent of each other (e.g., linearly de-correlated). Accordingly, in order to properly select the embedding delay $\tau$, estimates of the decorrelation time can be used. 

\begin{figure}
\centering
\resizebox{0.5\textwidth}{!}{\includegraphics*{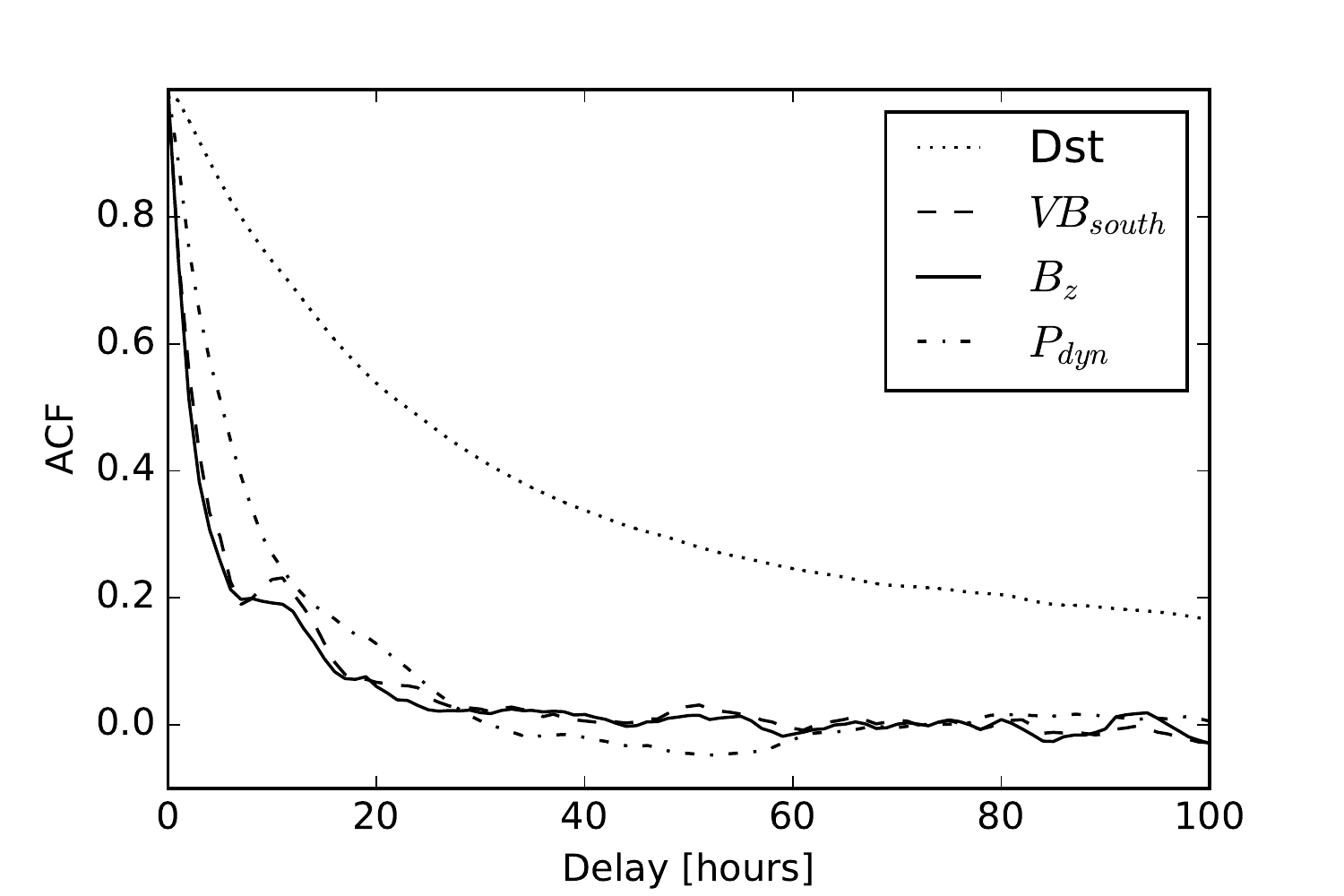}}
\caption{Autocorrelation functions (ACFs) for the time series of the geomagnetic activity index Dst and the solar wind parameters $VB_{south}$, $B_z$ and $P_{dyn}$ used to determine a suitable embedding delay $\tau$.}
\label{fig:embedding}
\end{figure}

Figure~\ref{fig:embedding} shows the autocorrelation functions of the four considered variables. Exploiting previous experience from other geoscientific applications \citep{Donges2011} that moderate variations of $\tau$ often have only a minor and mostly quantitative rather than qualitative effect on the results of subsequent analyses, we do not need to search for a distinct optimum value of $\tau$. Instead, we make use of the observation that for the entire one-year Dst index time series, the autocorrelation function decays within a few days to values below $1/e$. This motivates the choice of $\tau=100$ hours to be used in the remainder of the paper, which is in accordance with the previously reported long-range correlations in the Dst index emerging during geomagnetically active periods that imply a slow decay of serial dependencies in the corresponding temporal fluctuations \citep{Balasis2006,Donner2013}. Moreover, this value is consistent with the results presented by \cite{Johnson2005} who showed that strong nonlinear magnetospheric dependencies are statistically significant up to about one week. Further discussion on this choice is provided in some separate work \citep{dstpaper}. For the solar wind $VB_{South}$, $B_z$ and $P_{dyn}$, we consistently use $\tau=24$ hours to account for the comparably fast decay of autocorrelations in these observables. 

For the embedding dimension $m$, we use $m=3$ for all four variables as a trade-off between the possibly larger dimensionality of the observed fluctuations and the increased requirements for the length of the time series that needs to be considered when operating in higher embedding dimensions. Note that the latter are incompatible with the demand for the highest possible temporal resolution when studying dynamical complexity within running windows in time (see below for details).

\subsection{Recurrence plots}

Recurrence plots allow us to visualize the timing of observations of dynamically similar states of a system based on their mutual closeness in phase space \citep{Eckmann1987,Marwan2007}. In the case of the coupled solar wind--magnetosphere system, we use the embedded time series $\{X(t_i)\}_{i=1}^N$ as described above to define a binary recurrence matrix as
\begin{equation}
R_{ij}(\varepsilon)=\Theta(\varepsilon-\|X(t_i)-X(t_j)\|),
\end{equation}
\noindent
where $\Theta(\cdot)$ denotes the Heaviside function, $\|\cdot\|$ some norm in the phase space (in our case, we use the maximum norm, also known as $L_{\infty}$ or Chebyshev norm), and $\varepsilon$ the threshold distance used to define whether or not two embedded state vectors are close to each other. The graphical visualization of this matrix is known as the recurrence plot.

Different from the three solar wind-related variables, the Dst index takes only integer values, which would cause problems when evaluating the recurrence matrix with $\varepsilon$ chosen such that a dedicated recurrence rate $RR$ (i.e., the fraction of pairs of state vectors that are considered to be mutually close) is obtained. In order to avoid this effect, we add artificial Gaussian white noise with a standard deviation of $10^{-5}$ times the respective standard deviation to all considered time series. Different realizations of this noise have been found to have only negligible effects on the results described in the remainder of this paper (not shown). Note that the imposed noise can be thought of as representing the observational uncertainty and discretization error, and its actual variance will not affect the obtained results as long as it is small compared to the intrinsic discretization step of Dst.

In the following, we will exclusively consider the case of $RR=0.05$, which has been found to be a reasonable value for recurrence analysis in many examples of geoscientific time series \citep{Donges2011PNAS,Donges2015b}.

We emphasize that recurrence plots have already been used in the context of geomagnetic activity indices as well as related observables by \citet{Ponyavin2004, March2005, Dendy2006, Unnikrishnan2010}, however, mostly for visualization purposes. \citet{March2005b} used two index time series as examples to illustrate how to infer time-localized information on the mutual information from time series. Recently, \citet{Mendes2016} presented a first study aiming at obtaining quantitative information on high intensity and long duration continuous auroral activity from recurrence plots. In turn, regarding different solar activity indicators, a number of studies have utilized recurrence plots to characterize the underlying nonlinear dynamics \citep[see][and references therein]{Donner2008,Stangalini2017}.

\subsection{Recurrence quantification analysis (RQA) and recurrence network analysis (RNA)}

Beyond simple visual inspection of the recurrence matrix with the associated recurrence plots, a multitude of quantitative measures based on the pattern of recurrences can be used to reveal different aspects of a system's underlying  dynamical complexity. In this study, we employ three selected measures that have been found particularly suitable for this purpose when they were applied to time series from different fields of sciences (see \citet{dstpaper} for further details on their applications to the Dst index):

(1) In a recurrence plot, non-interrupted diagonal line structures formed by recurrent pairs of state vectors indicate that similar states tend to evolve similarly over a certain period of time. This property is captured by the mean diagonal line length that can also be interpreted as a measure of predictability \citep{Zbilut1992,Webber1994}. In the case of entirely random dynamics, diagonals only occur by chance and are most commonly short in length \citep{Marwan2007}. To distinguish between deterministic and stochastic dynamics, one can thus consider the so-called ``degree of determinism''
\begin{equation}
DET = \frac{\sum_{d=d_{min}}^{d_{max}} d\ p(d)}{\sum_{d=1}^{d_{max}} d\ p(d)}
\end{equation}
\noindent
as one of the most standard RQA measures, where $d$ denotes the length of a diagonal line, $p(d)$ is the associated probability density function, $d_{max}$ is the length of the longest diagonal (except for the main diagonal in the plot), and $d_{min}\geq 2$ (we use $d_{min}=2$ in this study to cover also cases where the maximum line length is relatively small). $DET$ gives the fraction of recurrences confined in diagonal structures and as such provides a heuristic measure that takes values close to one in the case of deterministic (predictable) dynamics, but lower values for stochastic (less predictable) behavior. However, note that values of $DET$ alone do not allow for identifying a possibly deterministic nature of a signal.

(2) In a similar way as diagonal line structures, non-interrupted vertical line structures formed by recurrent state pairs in a recurrence plot indicate that a system's state changes slowly with time \citep{Marwan2002}. With $p(v)$ being the probability density function of the vertical line length $v$, one convenient measure to quantify this aspect is the trapping time
\begin{equation}
TT = \frac{\sum_{v=v_{min}}^{v_{max}} v\ p(v)}{\sum_{v=v_{min}}^{v_{max}} p(v)}.
\end{equation}
\noindent
Low $TT$ values generally indicate fast changes of the system's state, whereas high values correspond to slow changes. Unlike $DET$, $TT$ is not normalized and can take any non-negative value between $0$ (when there are no vertical structures in the recurrence plot) and the length $N$ of the considered time series (constant time series). Consistent with $DET$, we will consider a minimum line length of $v_{min}=2$ in this study. In general, the use of (diagonal and vertical) line-based recurrence measures is referred to as recurrence quantification analysis (RQA) \citep{Marwan2007,Marwan2014}.

(3) Making use of the formal equivalence between the binary recurrence matrix $R_{ij}(\varepsilon)$ and the adjacency matrix $A_{ij}(\varepsilon)=R_{ij}(\varepsilon)-\delta_{ij}$ (with $\delta_{ij}$ being the Kronecker delta) of an undirected and unweighted network, it is possible to exploit the toolbox of complex network analysis to characterize different geometric properties of the system's organization in phase space \citep{Marwan2009,Donner2010,Donner2011}. In recent applications of recurrence network analysis (RNA) to artificial as well as real-world time series from various fields \citep{Marwan2009,Zou2010,Donges2011,Donges2011PNAS,Donges2015}, it has been found that the recurrence network transitivity
\begin{equation}
\mathcal{T}=\frac{\sum_{i,j,k} A_{ij}A_{ik}A_{jk}}{\sum_{i,j,k} A_{ij}A_{ik}}
\end{equation}
\noindent
provides a particularly useful measure for discriminating between qualitatively different types of dynamics. Specifically, this measure is closely related with the effective degrees of freedom of the system's dynamics and can be used to obtain an easily calculable generalization of a fractal dimension \citep{Donner2011EPJB,Donges2012}. Specifically, high (low) transitivity values indicate a low (high) dimensionality of the observed dynamics.

In addition to the aforementioned three recurrence measures, both RQA and RNA provide a variety of further characteristics that can be used for tracing different aspects of the dynamical complexity of the system under study. Here, we focus on just these three specific characteristics that have been demonstrated to be particularly useful in previous applications to different geoscientific time series \citep{Donges2011,Donges2011PNAS}.

\subsection{Sliding windows analysis}

In order to trace temporal changes in the dynamical complexity of the Earth's magnetosphere coupled to the solar wind, we do not solely attempt a global characterization of the system's state, but consider sliding windows in time. For this purpose, we construct recurrence plots of all four variables of interest with a fixed recurrence rate $RR=0.05$ and windows with a width of $w=168$ hours (7 days, covering the typical durations of geospace storms) and a mutual offset of $\Delta w=1$ hour. By conserving the recurrence rate $RR=0.05$, we ensure that the results obtained for all four variables and three recurrence measures are quantitatively comparable over time. 

As a reference time coordinate, we choose the second embedding coordinate, implying that information from past and future conditions is considered in a balanced way in the different recurrence characteristics \citep{dstpaper}. While such an approach is not particularly suited to the search for possible precursory structures in solar wind parameters and geomagnetic activity indices related to the initiation phase of magnetic storms, it allows the best possible differentiation between the dynamical characteristics of storm and non-storm periods \citep{dstpaper}. It should also be noted that due to the finite embedding delay $\tau$, all running windows effectively include information from time intervals $[t_{c}-w/2-\tau,t_{c}+w/2+\tau]$, with $t_{c}$ denoting the window midpoint. In other words, the effective time window of data utilized by our analysis has a width of $w+2\tau$, which is considerably larger than $w$, especially in the case of the Dst index ($\tau=100$ hours). This detail needs to be kept in mind when interpreting the results of our recurrence analysis, as well as when employing the proposed analysis approach to identify dynamical structures in the solar wind corresponding to possible precursory phenomena of magnetic storms.

The aforementioned considerations imply that we employ sliding windows over the three-dimensional state vectors of the reconstructed phase spaces of all four variables of interest instead of the original univariate time series. Thereby, we explicitly disregard the time-dependence of the temporal correlation structure of the variables of interest during different overall conditions of the system (which are particularly well known for the Dst index \citep{Balasis2006,Donner2013}) that might otherwise require using different embedding parameters ($\tau$,$m$) for different time windows. In turn, choosing the latter parameters adaptively for each window would, on the one hand, require much larger windows than the chosen width $w$ and, on the other hand, hamper the comparability of dynamical characteristics obtained for different windows due to the different time-scales covered by the respective embeddings.

\section{Results}\label{sec:performance}

\subsection{Recurrence plots of all variables}

\begin{figure*}
\centering
\resizebox{0.75\textwidth}{!}{\includegraphics*{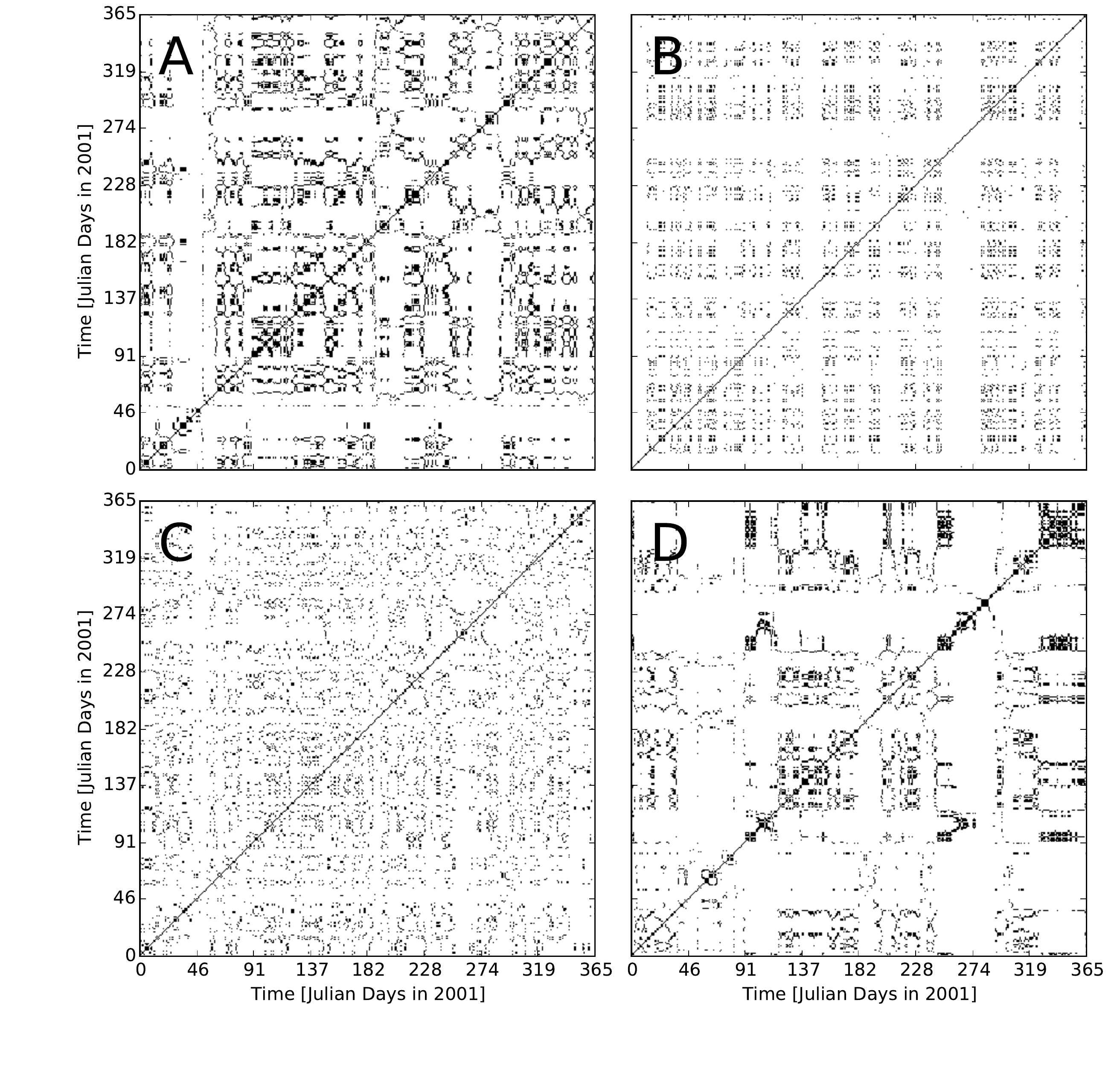}}
\caption{Recurrence plots for the time series of (A) the geomagnetic activity index Dst and the solar wind parameters (B) $VB_{south}$, (C) $B_z$ and (D) $P_{dyn}$ (see also Fig.~\ref{fig:data}), obtained with a constant global recurrence rate of $RR=0.05$.}
\label{fig:rps}
\end{figure*}

Figure~\ref{fig:rps} shows the recurrence plots for the full one-year time series of the Dst index and the three solar wind parameters analyzed in this work. As a first observation, we find that the results for the Dst index (Fig.~\ref{fig:rps}A) clearly indicate the periods of magnetospheric quiescence as areas characterized by a larger density of recurrences, whereas storm periods are marked by a larger range of values of the Dst index and, hence, fewer recurrences. In general, due to the significant changes in the amplitudes of fluctuations of most considered observables (except for $P_{dyn}$) along with geomagnetic activity periods as compared to their values during non-storm conditions, the local recurrence rate (density of recurrences in parts of the plot) varies markedly. In the sliding windows analysis presented below, we have instead fixed $RR$ individually for each window, implying that for the specific case of the Dst index, the window-specific recurrence threshold $\varepsilon$ takes larger values during storm than during non-storm periods in order to compensate for the overall larger variance (not shown).

Unlike for Dst and $P_{dyn}$, the recurrence plot of the solar wind parameter $B_z$ (Fig.~\ref{fig:rps}C) does not clearly discriminate between normal and disturbed periods of magnetospheric variability. In turn, the corresponding results for $VB_{south}$ reveal the presence of distinctive patterns only for the second storm period around the November 2001 magnetic storm (Fig.~\ref{fig:rps}B). 

Taken these observations together, a temporary loss of recurrences in this visual recurrence plot analysis is compatible with a larger variability (in the context of larger distances between state vectors in the reconstructed phase space) of the studied observable. As shown for the Dst index in previous studies  (see for instance the entropy-based studies by \citet{Balasis2008,Balasis2009}), this directly translates into a lower dynamical complexity (or a higher degree of organization of the system). On the other hand, the magnetosphere acts as a nonlinear filter to the temporal variations in the solar wind forcing, which can lead to similar magnetospheric dynamics during storm periods (as illustrated by similar Dst values during the two considered time windows with intense magnetic storms) even though the dynamical characteristics of the input variables (solar wind) can be distinctively different, as suggested by our results for $VB_{South}$. Even more, the temporal recurrence patterns of solar wind variables (input) and Dst index (output) can be qualitatively different. This finding is compatible with the hypothesis that during phases of an externally perturbed magnetosphere, additional internal processes are triggered and take place in the magnetosphere that lead to different dynamical complexity levels in input and output variables.

\subsection{Time-dependence of recurrence characteristics}

\begin{figure*}[th]
\centering
\resizebox{0.9\textwidth}{!}{\includegraphics*{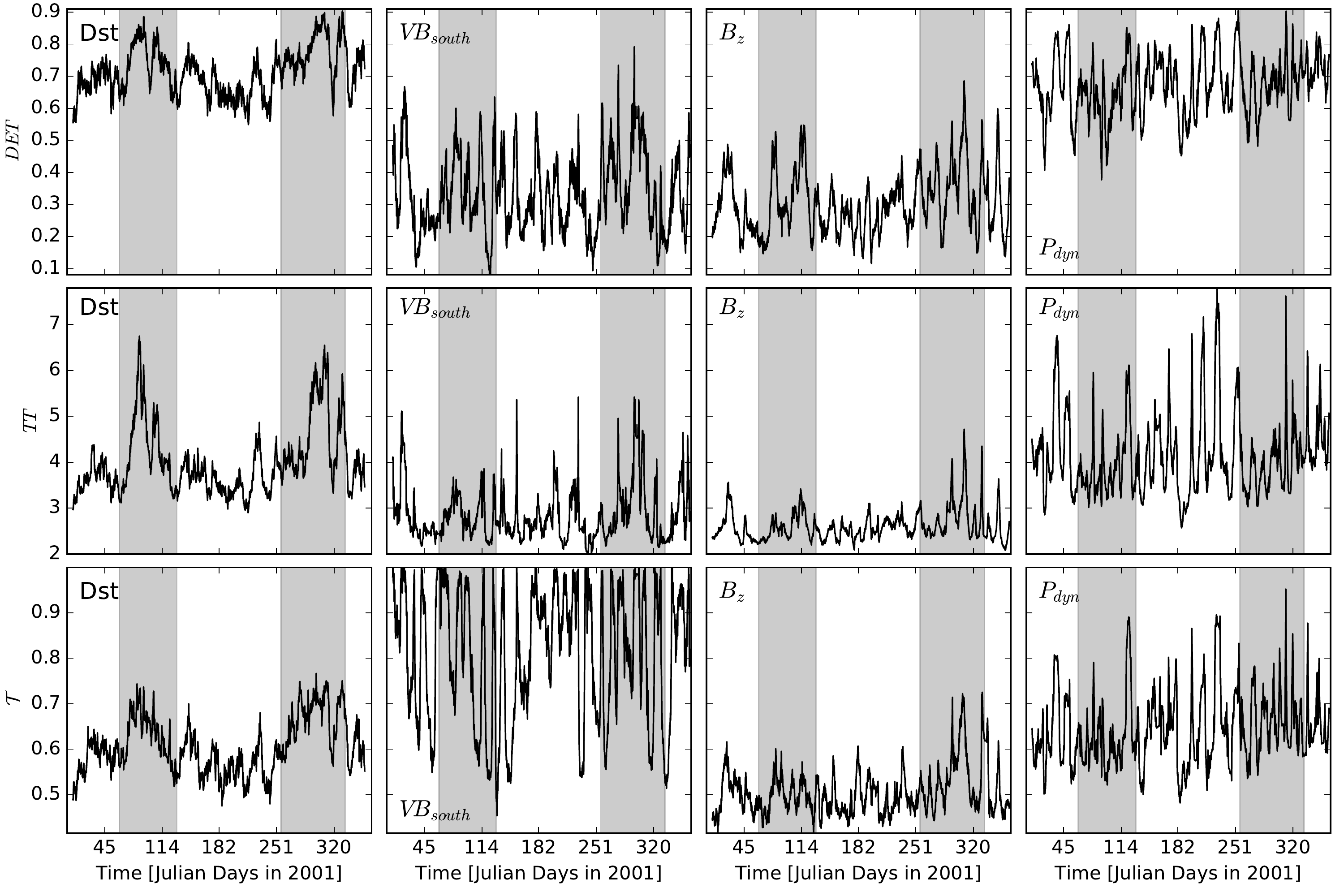}}
\caption{Temporal variations of $DET$, $TT$ and $\mathcal{T}$ (from top to bottom) calculated over sliding windows with a width $w$ of $168$ hours (one week) for the geomagnetic activity index Dst and the solar wind pa\-ra\-meters $VB_{South}$, $B_z$, $P_{dyn}$ (from left to right). Gray shaded areas indicate the two time intervals considered as storm periods, whereas the time intervals in between are considered as epochs of quiescence (see Section~\ref{sec:data} for details).}
\label{fig:sliding}
\end{figure*}

The results of our recurrence analysis for sliding windows in time are presented in Fig.~\ref{fig:sliding}. We clearly recognize that as expected from the qualitative recurrence plot characterization above, all three recurrence measures ($DET$, $TT$ and $\mathcal{T}$) exhibit marked variations with time, which are interpreted as changes in the dynamical complexity of different components of the coupled solar wind--magnetosphere system. Since both the solar wind and the magnetosphere exhibit strong fluctuations on a wide variety of timescales, there is no distinct baseline state for neither of the recurrence measures considered. Instead, they display variations on different timescales. In the following, we will discuss in some more detail (i) what information can be obtained from the typical ranges that the values of each recurrence measure take during the specific one year of observations, and (ii) how the respective recurrence characteristics differ between storm and quiescence periods as well as between the two studied storm periods.

\subsection{Typical complexity levels for all variables}

As a first step towards a more detailed interpretation of our recurrence analysis results, we compare the ranges of values that all three considered recurrence measures take for the four variables of interest. Here, we aim to characterize the typical complexity levels of the latter from different perspectives.

Regarding $DET$, Fig.~\ref{fig:sliding} demonstrates that for the Dst index, the estimated values range from about 0.53 to 0.91 and maxima appear during storm periods. These relatively high values indicate a moderate level of statistical predictability of the Dst index variations. In other words, if the temporal evolution of Dst index fluctuations in the past is known, it is possible to anticipate at least short-term trends in future variations for similar starting conditions -- a feature that is expressed in terms of diagonal line structures in the recurrence plot. Similar $DET$ values are observed for $P_{dyn}$ (with however much larger variance), whereas the two other solar wind parameters $VB_{South}$ and $B_z$ exhibit distinctively lower values with the calculated mean over all windows being approximately 0.3. This last observation suggests that (irregular) high-frequency variability is more pronounced in these two solar wind variables than in $P_{dyn}$ and the Dst index. This finding is consistent with the fact that the power spectral scaling exponent for the Dst index can take values both above 2 (during storm periods, indicating persistent behavior) and below 2 (during periods of quiescence, indicating anti-persistent dynamics), whereas it has been found to always stay below 2 for $VB_{South}$ even during periods of intense geomagnetic activity triggered by enhanced solar activity \citep{Balasis2006}.

The trapping time $TT$ of the Dst index variations takes a wide range of window-wise values between 3.0 and 6.8, with maximum values clearly coinciding with storm periods. For $P_{dyn}$, the obtained values of $TT$ cover an even slightly larger range. For the two other solar wind variables ($VB_{South}$ and $B_z$), both the minimum and maximum values of $TT$ are clearly smaller than for Dst and $P_{dyn}$, indicating again more (irregular) high-frequency variability and the absence of time periods during which the respective observable varies only weakly. This general distinction between the $TT$ ranges for the Dst index and $P_{dyn}$ versus $VB_{South}$ and $B_z$ is in agreement with the findings reported above for $DET$, indicating that the observed differences in the two RQA measures reflect dissimilar short-term fluctuations of the individual observables.

Finally, $\mathcal{T}$ provides a measure for the dimensionality of the time series \citep{Donner2011EPJB} (i.e.\ the redundancies among components of the embedding vectors). Notably, for $VB_{South}$, we observe numerous time intervals with $\mathcal{T}$ approaching this measure's limit value of 1 (indicating a zero-dimensional object in the underlying dynamical system's phase space, i.e., a fixed point). Because of this very specific behavior, we will not use $\mathcal{T}$ for further interpretation of the $VB_{South}$ records. In turn, the variability of $\mathcal{T}$ for Dst and the two other solar wind variables exhibits better interpretable features. Specifically, the Dst index and $P_{dyn}$ have larger average values of $\mathcal{T}$ than $B_z$, indicating again the presence of lower-dimensional dynamical structures, while $B_z$ appears to be ``more stochastic''. In general, $P_{dyn}$ shows the largest overall values of $\mathcal{T}$ (with the exception of $VB_{South}$ with its distinct behavior as described above) during a few distinct time intervals corresponding to both, storm and non-storm periods. We observe that these periods correspond to situations where $P_{dyn}$ drops and remains at relatively low values for a certain period of time (see also Fig.~\ref{fig:data}). This is consistent with the temporal profile of $\mathcal{T}$ previously observed for other geoscientific data sets \citep{Donges2011PNAS} and can be interpreted as an increased regularity of fluctuations in comparison with the ``typical'' values.

\subsection{Differences between storm and non-storm periods}

\begin{figure*}
\centering
\resizebox{0.75\textwidth}{!}{\includegraphics*{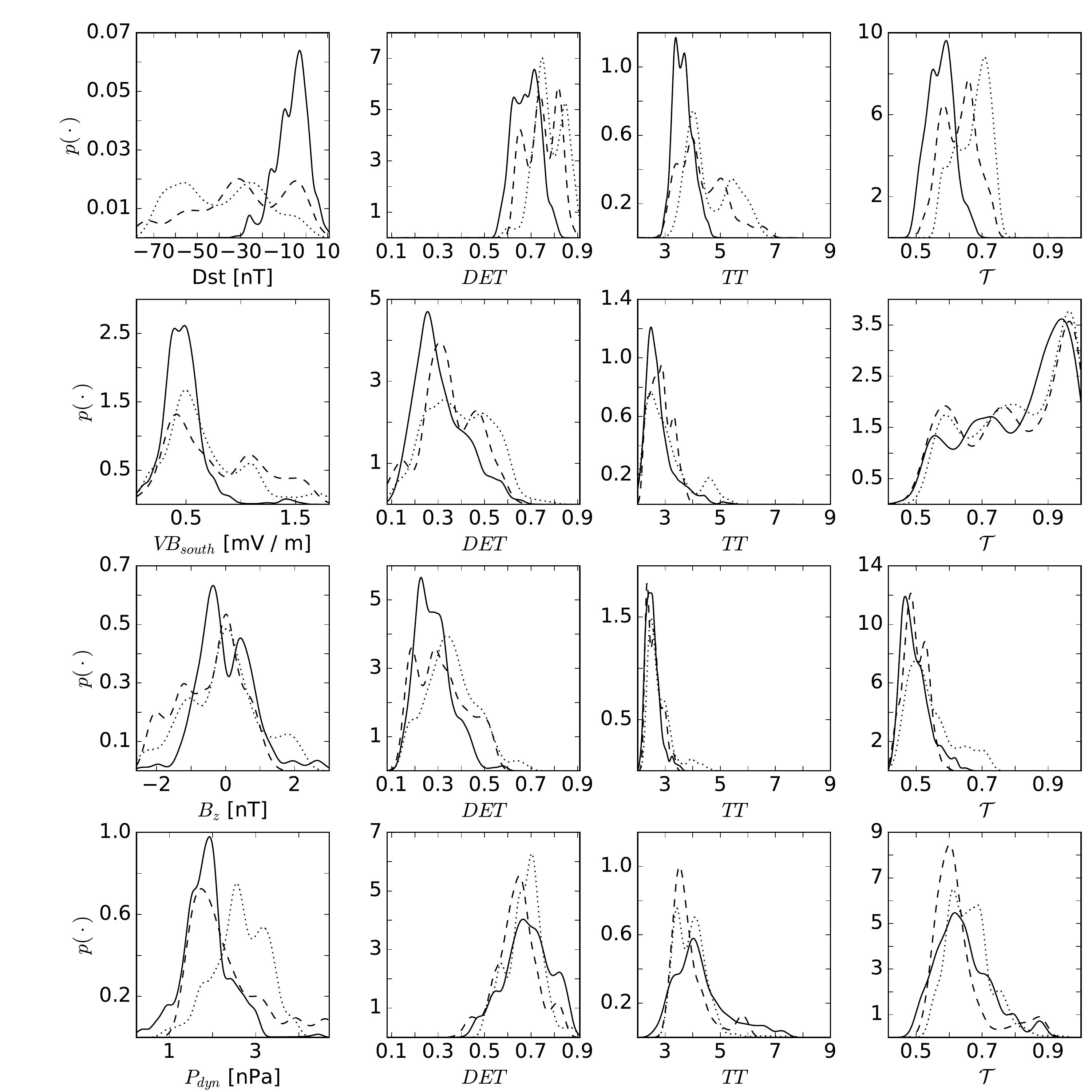}}
\caption{Probability density functions of the four considered magnetospheric and solar wind variables (from top to bottom: Dst index, $VB_{South}$, $B_z$ and $P_{dyn}$) and their associated recurrence characteristics (from left to right: window-wise mean value of each variable, $DET$, $TT$ and $\mathcal{T}$). Different line styles correspond to the three quiescence periods (solid), first storm period (dashed) and second storm period (dotted) as explained in the text.
}
\label{fig:scatter}
\end{figure*}

For the Dst index, the three considered recurrence measures clearly differentiate between storm and non-storm periods. In particular, $DET$, $TT$ and $\mathcal{T}$ reach higher values during periods of increased geomagnetic activity than in periods of quiescence. This finding is in agreement with the general observation that during periods of enhanced geomagnetic activity, the magnetosphere exhibits a larger degree of dynamical organization, which is also reflected in longer timescales of variability becoming more relevant (e.g.\ the initiation and recovery phases for sequences of magnetic storms). As mentioned above, it has been demonstrated previously that this behavior is expressed in terms of persistent dynamics, reduced dynamical disorder characterized by lower values of several entropy measures, and stronger autocorrelations \citep{Balasis2006,Balasis2008,Balasis2009,Balasis2011a,Balasis2011b,Donner2013, Balasis2013}. The aforementioned finding is further illustrated in Fig.~\ref{fig:scatter} (first row), which demonstrates that the distributions of values of $DET$, $TT$ and $\mathcal{T}$ differ from each other during storm and non-storm periods. As can be inferred from Fig.~\ref{fig:sliding} (first column), this difference visually appears most pronounced for $TT$. Note that while the difference in the distribution of recurrence characteristics is in most cases apparent from visual inspection, appropriate statistical testing against the null hypothesis of identical distributions (e.g., using Kolmogorov-Smirnov or similar statistics) is not straightforward since due to the strong overlap between successive time windows, the different values exhibit strong serial correlations and, thus, violate the common independence condition.

In turn, regarding the three solar wind observables, we find no similarly clear difference between the recurrence characteristics observed during storm and non-storm periods (Fig.\ref{fig:scatter}, second to fourth rows). In general, the three considered measures cover similar ranges of values during periods of increased geomagnetic activity and quiescence; however, all three recurrence characteristics for $B_z$ differ in the maximum values reached during storm periods. A similar conclusion cannot be drawn for $VB_{South}$ and $P_{dyn}$, where we observe an absence of marked differences in the recurrence characteristics during storm versus quiescence phases of magnetospheric dynamics. This points to a nonlinear and highly context-specific response of the magnetosphere to temporal changes in the dynamical characteristics of the solar wind variables, where similar dynamical complexity in the interplanetary medium may correspond to different overall conditions resulting in different levels of magnetospheric activity.

Specifically, a fast moving magnetic cloud with organized internal magnetic field is expected to cause an intense magnetic storm if it engulfs Earth, but this condition is not sufficient. A magnetic cloud (or, likewise, a CME triggering a magnetic storm) should have a magnetic field with a significantly negative $B_z$ component for reconnection to occur at the dayside magnetopause. \citet{Voros2005b} have shown that, in addition to the geo-effective southward component of IMF, intermittency, small-scale rapid changes, singularities and non-Gaussian statistics of IMF fluctuations play an important role in the solar wind--magnetosphere interaction. Figure ~\ref{fig:sliding} shows that for $B_z$, $DET$ and $\mathcal{T}$ exhibit a generally similar time evolution as the corresponding recurrence characteristics for the Dst index (but at a clearly lower level of the respective measures). This underlines that geomagnetic fluctuations at the typical timescales of geospace storms result from the simultaneous action of multiple magnetospheric processes, including the component represented by intermittent solar wind fluctuations.	

Generally, CMEs behave as nonlinearly interacting dynamical systems characterized by the presence of large-scale coherent structures of different sizes that decrease the degree of multifractality in the IMF's $B_z$ component because they privilege only a few scales \citep{Bolzan2012}. After the passage of a CME, an increase in the complexity is necessary to promote the dissipation of energy. However, some CMEs, like the first event in March 2001 studied in this work, are characterized by slow solar wind (approximately 550 km$\cdot$s$^{-1}$), which is more intermittent than fast solar wind and contributes to the intermittency of the interplanetary medium \citep{Bruno2003}. This is reflected in the lower maximum values of the three recurrence measures during this specific time period as compared to the second event in November 2001. However, even in the case of fast solar wind, intermittency increases with the heliocentric distance until it reaches the Earth's orbit, and IMF fluctuations tend to be more intermittent than velocity fluctuations \citep{Bruno2003}. 

In turn, fluctuations of the interplanetary electric field as reflected by $VB_{South}$ are considered responsible for initiating magnetospheric substorms, while its quasi-steady component plays the central role in the enhancement of the ring current that is monitored by the Dst index  \citep{Kamide2001}. This is also corroborated by previous findings of \citet{Balasis2006}, who have observed this behavior in geomagnetic time series where the degree of dynamical complexity of the Dst index is reduced in the presence of powerful oscillations. The lack of similarity between the recurrence characteristics of $VB_{South}$ and the Dst index in Fig.~\ref{fig:sliding} also suggests that only a part of the Dst index variation can be explained as a direct response to $VB_{South}$. In the past, \citet{Price1993} focused on an interval when the IMF had a nearly constant $B_z$ component to be able to find some evidence for a deterministic non-linear coupling between the solar wind forcing, expressed by the $VB_{South}$ electric field component, and the terrestrial magnetosphere. Besides the IMF and electric fields, a whole set of variables, including solar wind velocity, density, $P_{dyn}$ and plasma $\beta$, affect to different degrees the energy input to the magnetosphere. This is reflected by the recurrence-based characteristics of the selected solar wind parameters considered in this study.

\subsection{Differences between the two storm periods}

The magnetic storms of 31 March and 6 November 2001 were among the 11 superintense storms (Dst $\leq$ -250~nT) that occurred during solar cycle 23 \citep{Echer}. The first event was caused by a combination of sheath and magnetic cloud fields while the second one by sheath fields alone \citep{Echer}. This is reflected in the $TT$ values of the solar wind variables and in the response of the magnetosphere as expressed by the Dst index. 

In the gray shaded area of Fig.~\ref{fig:sliding} that is related to the second magnetic storm period, $TT$ exhibits a pronounced maximum for both solar wind parameters $B_z$ and $P_{dyn}$, while there are no peaks of similar magnitude during the first magnetic storm period. Specifically, the peak values of all three recurrence measures for $P_{dyn}$ are observed between days 285 and 332.5 of the year 2001, and their timing of occurrence corresponds to the onset phase of the 6 November 2001 magnetic storm. This finding is in accordance with \citet{Wang2003}, who argued that the ring current injection increases when the magnetosphere is compressed by a particularly strong solar wind forcing and that the injection rate is proportional to $P_{dyn}$. We note that the Dst variations of the  6 November 2001 magnetic storm were also found to obey a power law with log-periodic oscillations \citep{Balasis2011c}, which is a sign for the emergence of discrete scale invariance in the magnetosphere. 

One possible interpretation of our results on the values of $TT$ is as follows. During the first storm phase, $VB_{South}$ and $B_z$ exhibit faster variations (lower $TT$) than Dst and $P_{dyn}$ that both show comparable maximum values (around 6) reflecting similar time scales at which the two variables change. During the second storm phase, we find a general tendency towards higher $TT$ values than during the first storm phase (and, hence, slower changes) for all three solar wind variables. In order to explain this observation, we suggest that in the first case, the magnetosphere (and in particular the ring current) has absorbed or screened the faster changes of the two solar wind variables $VB_{South}$ and $B_z$ and follows more closely the (somewhat slower) variations of $P_{dyn}$. This situation is compatible with the scenario proposed by \citet{Wang2003}), who argued that the ring current injection increases when the magnetosphere is more compressed by a particularly strong solar wind forcing and that the injection rate is proportional to $P_{dyn}$. In contrast, during the second storm phase, the ring current seems to have more closely followed the variations in $VB_{South}$ and $B_z$.


\section{Discussion} \label{sec:discussion}

Based on the results described in the previous section, we suggest that recurrence based complexity measures have a great potential to trace temporal variations in the dynamical complexity of geomagnetic and solar wind dynamics, but also other nonstationary geophysical observables. In particular, the dynamical complexity profile of magnetospheric fluctuations during storm and non-storm conditions (which we have studied in terms of $DET$, $TT$ and $\mathcal{T}$, capturing the regularity of fluctuations in the Dst index from different perspectives) is in good agreement with the existing body of literature on this subject (cf.~\citet{Balasis2009}), as will be discussed in the following.

\citet{Consolini2008} investigated long-term variations in the dynamical state of the Earth's magnetosphere in terms of the Dst index. Their results clearly demonstrated the non-equilibrium nature of magnetospheric dynamics. Specifically, the Earth's magnetosphere behaves like expected for a system far from equilibrium due to the continuous interaction with the time-dependent solar wind forcing. The presence of two different dynamical regimes -- near and away from a non-equilibrium stationary state -- has been independently confirmed by other studies. Specifically, \citet{Sitnov2001} provided evidence that substorms exhibit dynamical characteristics that are typical for phase transitions. This picture is consistent with the findings of \citet{Balasis2006} who reported the transition from anti-persistent to persistent behavior when an intense magnetic storm is imminent. Moreover, the metastability and topological complexity of the geomagnetic variations established with the model of \citet{Chang1999} are in good agreement with the transitions from the observed pre-storm activity to magnetic storms that have been found in our study. \citet{Chang2003,Chang2004} and \citet{Voros2005a} provided indications for the presence of intermittent turbulence in space plasmas, which further supports our results. Furthermore, the statistical properties of magnetic fluctuations in the Venusian magnetosphere determined by \citet{Voros2008} point to multi-scale turbulence at the magnetosheath boundary layer and near the quasi-parallel bow shock.

In addition, a reduction of multiscale complexity was observed in the geomagnetic activity at high latitudes before strong substorms. With the use of cellular automata models, \citet{Uritsky1998} and \citet{Uritsky2001} demonstrated transitions between sub-critical, critical and super-critical states. A similar behavior was found in the spatial scaling of the auroral brightness \citep{Uritsky2006,Uritsky2008}. The multiscale complexity of geomagnetic substorms was explored in a series of studies \citep{Chang1992,Consolini1997,Chapman1998,Klimas2000,Lui2000}, while the first discussions on the critical nature (self organized criticality - SOC) of geomagnetic storms were initiated by \citet{Consolini1997}, \citet{Uritsky1998} and \citet{Chapman1998}. \citet{Wanliss2004} and \citet{Wanliss2010} reported evidence for intermittency and non-Gaussianity associated with large magnetic storms using symbolic dynamics analysis of the Dst time series. These findings strengthen the hypothesis of a constantly out-of-equilibrium ring current that undergoes state changes in terms of multiplicative cascades. In general, the results obtained in this study further support the existence of two distinct states of the magnetosphere corresponding to storm and non-storm conditions.

Attempts to determine the non-linear properties of the magnetospheric system from just its response -- no matter how robust and complete the diagnostic means are -- would not be meaningful without considering the solar wind forcing. Following the observed power law behavior of the AE index and the southward component of the IMF, the multiscale properties of magnetospheric dynamics have been interpreted in terms of intermittent turbulence and SOC \citep{Consolini1997}. In the seminal paper by \citet{Tsur1990}, the magnetosphere was considered as an input--output system to find a typical timescale of 5 hours allowing to disentangle the internal fast and bursty dynamics of the magnetosphere from the directly driven one. In this study, we analyzed the response of the magnetosphere as expressed by the Dst index while the solar wind input has been represented by the IMF's $B_z$ and the dawn--dusk component of the electric field $VB_{South}$ together with $P_{dyn}$ to provide evidence that (at least) a part of the Dst index variations can be explained as a direct nonlinear response of the magnetosphere to the solar wind. 

The consideration of multiple measures from RQA and RNA -- all based on the same underlying structure of the corresponding recurrence plots, but generally characterizing different aspects of dynamical complexity -- allowed us to identify statistics that are particularly suited for distinguishing periods of magnetic storms and quiescence based on the dynamical complexity of Dst index and solar wind parameter fluctuations. Specifically, there are considerable small-scale differences between the recurrence plots of the Dst index and the three solar wind parameters, IMF $B_z$, $VB_{South}$ and $P_{dyn}$. Moreover, the large-scale structure of the recurrene plots is distinctively different for Dst and its potential drivers. In particular, the recurrence characteristics for the IMF $B_z$ exhibit rather distinct behaviors during the transition from periods of relative quiescence to magnetic storms, revealing the nonlinear response of the terrestrial magnetosphere to the geoeffective southward component of the IMF as CMEs or magnetic clouds approach the Earth's neighborhood.  

In the past, \citet{March2005} employed recurrence plots to visualize nonlinear correlations between the AE index and $VB_{South}$ times series. While they also observed many differences between the obtained recurrence plots at small scales that were attributed to fast fluctuations, the overall large-scale structure displayed by both variables was qualitatively similar. It seemed that the shared structure on the recurrence plots covering periods of both high and low activity is the result of electric fields in the solar wind related to effects on Earth observed on a timescale of the order of hours. Even though our analysis has not yielded any discernible change in the degree of determinism $DET$ during magnetic storms for $VB_{South}$ time series, a multiple-measure approach promises to provide interesting information from the study of the Earth's magnetic field variations, even below the general storm/quiescence variability. This perspective shall be further explored in future work, thereby extending this study to other geomagnetic activity indices employed for tracing magnetospheric phenomena like substorms.



\section{Conclusions} \label{sec:conclusion}

We have applied a set of three complementary recurrence based measures to study the temporal changes in dynamical complexity exhibited by the Dst index together with three characteristic variables of the solar wind during one year of observations near a solar maximum. Our results demonstrate that in the case of Dst, all three measures are able to trace variations associated with the time-dependent dynamical complexity of magnetospheric variability during a succession of storm and non-storm periods. Nonetheless, the measures exhibit different degrees of sensitivity with respect to such changing conditions resulting from characteristic structures like CMEs and magnetic clouds imprinted in the solar wind's dynamical properties. Specifically, the degree of determinism $DET$ takes relatively large values for the Dst index, the IMF $B_z$ and the $VB_{South}$ during storm time periods, but typically much lower values during periods of quiescence, indicating a more stochastic behavior. Low values of the trapping time $TT$ indicate fast changes of the system's state, whereas high values observed in the Dst index, the IMF $B_z$ and $P_{dyn}$ point towards slower changes, which are characteristic for a decrease in the complexity of the solar wind driver of magnetospheric disturbances. The recurrence network transitivity $\mathcal{T}$ can be used to define an easily calculable generalization of a fractal dimension and reaches higher values (indicating lower dimensionality) for both, the Dst index and $B_z$. 

Our results, together with other recent findings characterizing the multifractality of the interplanetary plasma, suggest that measures associated with recurrence plots provide valuable insights into the temporal structure of solar wind measurements with geomagnetic effects observed through the Dst index. The recurrence based complexity measures employed here serve as tools to detect solar wind structures leading to intense magnetospheric events, especially when the solar--terrestrial interaction is studied within the context of space weather forecasting.

We emphasize that the quantitative differences in the considered measures applied to different variables are clearly affected by the spectral properties of the respective time series. Specifically, for signals with a strong high-frequency content and hardly any variability at lower frequencies, large-scale structures in the recurrence plot are unlikely to emerge, and this would be reflected by relatively low values of our complexity measures. In turn, signals with a low level of high-frequency variability are prone to highlight long-term changes, which would result in large-scale recurrence patterns and, consequently, a higher probability of elevated values of the three considered recurrence characteristics. Future work should address this aspect more explicitly, e.g. by considering sophisticated filtering of the selected observables to retain only the variations below a certain minimum frequency. 

Finally, it is arguable that solar wind forcing and possible magnetospheric response are characterized by different temporal variability scales, which have not yet been accounted for in our present analysis (in particular, by keeping the embedding delay as a fundamental parameter of our analysis intentionally fixed at the same value during storm time and quiescence periods). To shed further light on the relevance of different scales, it might be advisable to disentangle variability at different scales and consider the dynamical complexity associated with fluctuations at individual timescales (as seen by recurrence properties or other nonlinear characteristics) independently. Even more, studying the possible transfer of information on the fluctuation properties across different scales might provide another fruitful subject of future studies.


%
%
%
%
%
%
%

\begin{acknowledgments}
This work has been financially supported by the bilateral Greek--German exchange project ``Transdisciplinary assessment of dynamical complexity in magnetosphere and climate: A unified description of the nonlinear dynamics across extreme events''  jointly funded by IKY and DAAD. Individual financial support of the authors has been granted by the Marie Curie Initial Training Network (ITN) program (FP7-PEOPLE-2011-ITN) under grant agreement no. 289447 for the ``Learning about Interacting Networks in Climate (LINC)'' project, the German Federal Ministry for Science and Education under grant agreement no. 01LN1306 for the BMBF Young Investigator’s Group CoSy-CC$^2$, the Government of the Russian Federation (funding agreement 14.Z50.31.0033 with the Institute of Applied Physics of RAS), and the International Research Training Group IRTG 1740/TRP 2014/50151-0, jointly funded by the German Research Foundation (DFG, Deutsche Forschungsgemeinschaft) and the S\~{a}o Paulo Research Foundation (FAPESP, Funda\c{c}\~{a}o de Amparo \`a Pesquisa do Estado de S\~{a}o Paulo). Numerical codes used for estimating RQA and RNA properties can be found in the software package \texttt{pyunicorn} \citep{Donges2015b}, which is available at \url{https://github.com/pik-copan/pyunicorn}. For the gap-filling of solar wind time series, we have utilized the R package \texttt{spectral.methods} developed by Jannis von Buttlar. Finally, the authors wish to express their special thanks to three anonymous reviewers of a topically related manuscript, whose detailed comments have provided essential suggestions leading to the work presented here.
\end{acknowledgments}

\end{document}